\documentclass{ws-procs9x6}

\begin{document}

\title{
A Method to Control Order of Phase Transition:\\
Invisible States in Discrete Spin Models
}

\author{Ryo Tamura}

\address{
Institute for Solid State Physics, University of Tokyo,\\
5-1-5, Kashiwanoha, Kashiwa-shi, Chiba, Japan, 277-8581
}

\author{Shu Tanaka}

\address{
Department of Chemistry, University of Tokyo,\\
7-3-1, Hongo, Bunkyo-ku, Tokyo, Japan, 113-0033
}

\author{Naoki Kawashima}

\address{
Institute for Solid State Physics, University of Tokyo,\\
5-1-5, Kashiwanoha, Kashiwa-shi, Chiba, Japan, 277-8581
}

\begin{abstract}
It is an important topic to investigate nature of the phase transition in wide area of science such as statistical physics, materials science, and computational science.
Recently it has been reported the efficiency of quantum adiabatic evolution/quantum annealing for systems which exhibit a phase transition, and we cannot obtain a good solution in such systems.
Thus, to control the nature of the phase transition has been also attracted attention in quantum information science.
In this paper we review nature of the phase transition and how to control the order of the phase transition.
We take the Ising model, the standard Potts model, the Blume-Capel model, the Wajnflasz-Pick model, and the Potts model with invisible states for instance.
Until now there is no general method to avoid the difficulty of annealing method in systems which exhibit a phase transition.
It is a challenging problem to propose a method how to erase or how to control the nature of the phase transition in the target system.
\end{abstract}


\bodymatter

\section{Introduction}

To study nature of the phase transition has been attracted attention in wide area of science such as materials science and statistical physics.
If we change control parameters such as temperature, pressure, and external field, we can observe phase transition in many materials.
In respective materials, a macroscopic ordered state ({\it e.g.} ferromagnetism, ferroelectoricity, superfluidity, and superconductivity) appears, which is characterized by each corresponding order parameter.
The Ising model was proposed in 1925, which is regarded as the most fundamental magnetic model to study the nature of the phase transition from a microscopic viewpoint in statistical physics\cite{RTIsing-1925}.
It is not too much to say that the Ising model triggers the development of statistical physics.
Since the Ising model can be easily generalized because of simplicity, a number of generalized models have been proposed up to the present.
The relation between the universality class and the symmetry which breaks at the transition point has been studied by analysis of these models.
In general, the Hamiltonian of the Ising model with site-dependent external magnetic field $h_i$ is expressed by
\begin{eqnarray}
 \label{RTeq:Ising}
 {\cal H}_{\rm Ising} = -\sum_{i,j} J_{ij} \sigma_i^z \sigma_j^z - \sum_i h_i \sigma_i^z,
  \quad
  \sigma_i^z = +1, -1,
\end{eqnarray}
where $\sigma_i^z$ is a microscopic variable at the $i$-th site.
Hereafter the $g$-factor and the Bohr magneton $\mu_{\rm B}$ are set to be unity for simplicity.
When $J_{ij}>0$, the interaction between the $i$-th and $j$-th sites is ferromagnetic, whereas for $J_{ij}<0$, the interaction is antiferromagnetic.
We can adopt the Ising model for not only analysis of phase transition observed in real materials but also information science/technology.
In information science/technology, a binary representation is a basic language.
Then, an interdisciplinary science which is the interface between statistical physics and information science has been developed in terms of the Ising model.
Actually some difficulties of information science/engineering were solved from a viewpoint of statistical physics\cite{RTNishimori-2001}.

Optimization problem is a problem to find the minimum/maximum value of real-valued cost/gain function.
Then, to solve an optimization problem corresponds to find the equilibrium state at finite temperature or the ground state of the Hamiltonian which expresses the target optimization problem.
In many cases we can represent optimization problem by the Ising model or its generalized model.
An algorithm which can solve optimization problem in a general way was proposed by Kirkpatrick.
This method is called simulated annealing\cite{RTKirkpatrick-1983,RTKirkpatrick-1984}.
In general, energy landscape of optimization problems is complicated as random spin systems\cite{RTMezard-1987,RTFischer-1993,RTYoung-1998}.
In the simulated annealing, we gradually decrease temperature and can obtain a not so bad or the best solution.
A characteristic transition time at the temperature $T$ is expressed $\tau \propto {\rm e}^{\beta\Delta E}$, where $\beta$ denotes the inverse temperature ($\beta=T^{-1}$) and $\Delta E$ represents a characteristic energy difference.
For simplicity, the Boltzmann constant $k_{\rm B}$ is set to be unity.
Since the probability distribution of equilibrium state is almost flat at high temperature, $\tau$ becomes short.
It was shown that by decreasing temperature slow enough, we can obtain the best solution of optimization problems by Geman and Geman\cite{RTGeman-1984}.
Since the simulated annealing is easy to implement, this is adopted in many cases\cite{RTLaarhoven-1987}.

The simulated annealing finds a not so bad solution or the best solution by making use of thermal fluctuation.
In 1998, on the contrary, Kadowaki and Nishimori proposed an alternative method of simulated annealing, which is called quantum annealing\cite{RTKadowaki-1998}.
In the quantum annealing, we decrease the quantum field ({\it i.e.} quantum fluctuation) instead of temperature in the simulated annealing\cite{RTBrooke-1999,RTFarhi-2001,RTSantoro-2002,RTSuzuki-2005,RTDas-2005}.
By using the quantum annealing, we can succeed to obtain a better solution than the solution obtained by the simulated annealing in many cases\cite{RTMatronak-2004,RTBattaglia-2005,RTTanaka-2007,RTKurihara-2009,RTSato-2009,RTMorita-2009,RTInoue-2010}.
Thus, the quantum annealing is expected as a powerful tool for optimization problems\cite{RTDas-2008,RTTanaka-2009,RTTanaka-2010,RTTanaka-2010book,RTChandra-2010,RTTanaka-2011a,RTTanaka-2011b,RTTanaka-2011c,RTOhzeki-2011}.
In order to improve this method more efficient, the quantum annealing from a viewpoint of statistical physics has been studied\cite{RTOhzeki-2010}.
Annealing methods which are based on statistical physics seem to work well in all optimization problems since these methods are easily performed.
However there is a weak point in both simulated annealing and quantum annealing.
When we decrease temperature/quantum field across a phase transition point, we obtain not so good solution in general\cite{RTDas-2005}.
Especially, if a first-order phase transition occurs during annealing, we can not obtain the best solution definitely\cite{RTYoung-2010}.
The situation is improved if a second-order phase transition takes place but some kind of critical slowing down exists\cite{RTSuzuki-2010}.
Thus, it is an important issue to investigate how to erase the phase transition or how to change the order of the phase transition from first-order to second-order.

In this paper we focus on how to control the order of the phase transition.
The organization of the rest of the paper is as follows.
In Section 2, we explain nature of the phase transition taking as examples some fundamental models.
In Section 3, we review how to change the order of the phase transition for preceding studies.
We take the Blume-Capel model and the Wajnflasz-Pick model for instance.
In Section 4, we consider the Potts model with invisible states.
In Section 5, we summarize this paper and show future perspective.

\section{Nature of the Phase Transition}

Phase transition can be categorized into two types according to singularity of the free energy as a function of control parameter in the thermodynamic limit.
If there is a singularity in differential coefficient of first order of free energy, a first-order phase transition occurs and then, energy and order parameter jump at the transition point.
From this feature,
a first-order phase transition is called discontinuous phase transition.
On the other hand, if there is a singularity in differential coefficient of second order of free energy, a second-order phase transition takes place.
In this case, energy and order parameter are continuous even at the transition point.
Then, a second-order phase transition is called continuous phase transition as against discontinuous phase transition.
When a second-order phase transition occurs, physical quantities near the transition point should be represented by power functions.
The exponents of these functions are called critical exponents.
A set of values of critical exponents corresponds to a universality class.
Universality class has been investigated exhaustively by analytical methods and numerical methods such as Monte Carlo simulation.
We can categorize a phase transition according to ``{\it encyclopedia} of universality class''
\footnote{
In some cases, novel universality class is found.
It should be noted that to explore new universality class itself is an important topic in statistical physics.
}.

\begin{figure}[t]
 \begin{center}
  \psfig{file=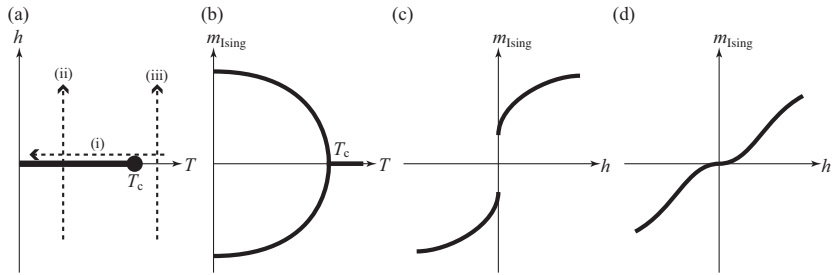,scale=0.8}
 \end{center}
 \caption{
 (a) Phase diagram of the ferromagnetic Ising model for $d\ge 2$.
 The bold line represents ferromagnetic phase, and the circle denotes the critical point $T_{\rm c}$.
 (b) Behavior of magnetization when we decrease temperature at $h=0$ as the line (i) in (a).
 In actual, either the upper or lower branch is selected.
 This phenomenon is called spontaneous symmetry breaking.
 A second-order phase transition occurs at $T_{\rm c}$.
 (c) Behavior of magnetization when we change external magnetic field at $T < T_{\rm c}$ as the line (ii) in (a).
 A first-order phase transition occurs.
 (d) Behavior of magnetization when we change external magnetic field at $T > T_{\rm c}$ as the line (iii) in (a).
 No phase transition happens.
 }
 \label{RTfig:IsingPT}
\end{figure}

Here we explain the phase transition of the ferromagnetic Ising model with nearest-neighbor interaction under homogeneous external magnetic field.
The Hamiltonian is given as
\begin{eqnarray}
 {\cal H}_{\rm Ising}^{\rm (2)} = 
  -J \sum_{\langle i,j \rangle} \sigma_i^z \sigma_j^z - h \sum_i \sigma_i^z,
  \quad (J > 0),
  \quad
  \sigma_i^z = +1, -1.
\end{eqnarray}
Hereafter $\langle i,j \rangle$ denotes the nearest neighbor spin pair on the defined $d$-dimensional lattice, and we take $J$ as the energy unit throughout this paper.
The phase diagram of this model for $d\ge 2$ is depicted in Fig.~\ref{RTfig:IsingPT} (a)\footnote{
On the one-dimensional lattice, there is no phase transition.
}.
The horizontal and vertical axes in this figure are temperature $T$ and external magnetic field $h$, respectively.
The bold line and the circle in Fig.~\ref{RTfig:IsingPT} (a) are the ferromagnetic phase and the critical point $T_{\rm c}$.
When we decrease temperature under zero field (see the line (i) in Fig.~\ref{RTfig:IsingPT} (a)), a second-order phase transition with spontaneous twofold symmetry breaking occurs at the critical point $T_{\rm c}$.
In this model, the order parameter is magnetization defined as
\begin{eqnarray}
 m_{\rm Ising} = \frac{1}{N} \sum_i \sigma_i^z,
\end{eqnarray}
where $N$ is the number of spins.
When the external magnetic field is zero, the behavior of magnetization is shown in Fig.~\ref{RTfig:IsingPT} (b).
On the other hand, when we change the external magnetic field with fixed temperature below $T_{\rm c}$ (see the line (ii) in Fig.~\ref{RTfig:IsingPT} (a)), a first-order phase transition occurs and the magnetization jumps at $h=0$ as depicted in Fig.~\ref{RTfig:IsingPT} (c).
When we change the external magnetic field with fixed temperature above $T_{\rm c}$ (see the line (iii) in Fig.~\ref{RTfig:IsingPT} (a)), the magnetization behaves as a smooth function shown in Fig.~\ref{RTfig:IsingPT} (d).
When we sweep external magnetic field along the line (ii) in Fig.~\ref{RTfig:IsingPT} (a) at finite speed, hysteresis curve is often observed because of existence of the metastable state.

Suppose we perform simulated/quantum annealing for systems which exhibit a first-order phase transition.
When we decrease temperature/quantum field across the transition point, the state is trapped in the metastable states.
Then we cannot obtain not so bad solution by annealing methods.
Next we consider the case for second-order phase transition.
Physical quantities converge to the equilibrium value slowly because of some kind of critical slowing down.
This nature was studied recently, which is called the Kibble-Zurek mechanism\cite{RTSuzuki-2010,RTKibble-1976,RTKibble-1980,RTZurek-1985,RTZurek-1993}.

The relation between the order of the phase transition and the symmetry which breaks at the transition point has been considered by using the standard ferromagnetic Potts model\cite{RTPotts-1952,RTWu-1982}.
The Hamiltonian of the ferromagnetic $q$-state Potts model is given as
\begin{eqnarray}
 {\cal H}_{\rm Potts} = -J \sum_{\langle i,j \rangle} \delta_{s_i,s_j},
  \quad (J>0),
  \quad
  s_i = 1, \cdots, q.
\end{eqnarray}
Since the $2$-state Potts model is equivalent to the Ising model, the $q$-state Potts model is regarded as a straightforward extension of the Ising model.
The Potts model is used for analysis of coloring problems and clustering problems\cite{RTKurihara-2009,RTSato-2009}, and plays an important role for not only statistical physics but also information science.
In this model on one-dimensional lattice, there is no phase transition at finite temperature for arbitrary $q$ as well as the Ising model.

It is convenient to introduce another representation of Kronecker's delta as 
\begin{eqnarray}
 \delta_{\mu,\nu} = \frac{1+(q-1){\bf e}^\mu \cdot {\bf e}^\nu}{q},
\end{eqnarray}
where ${\bf e}^\mu$ ($\mu=1,\cdots,q$) represents $q$ unit vectors pointing in the $q$ symmetric direction of a hyper-tetrahedron in $q-1$ dimensions\cite{RTWu-1982}.
Then the order parameter of this model can be defined as
\begin{eqnarray}
 \label{RTeq:ORDPotts}
 {\bf m}_{\rm Potts} = \frac{1}{N} \sum_i {\bf e}^{s_i}.
\end{eqnarray}
This model on two-dimensional lattice exhibits a second-order phase transition with $q$-fold symmetry breaking for $q\le 4$ whereas a first-order phase transition with $q$-fold symmetry breaking for $q >4$.
In a similar way, the relation between the order of the phase transition and the symmetry which breaks at the transition point for $d>2$ is also investigated as shown in Table \ref{RTtable:sPotts}\cite{RTKihara-1954}.

\begin{table}[t]
\tbl{Relation between the order of the phase transition and $q$ in the standard ferromagnetic Potts model on $d$-dimensional lattice. Note that $q$-fold symmetry breaks at the transition point.}
{
\begin{tabular}{@{}ccc@{}}\toprule
 Dimension $d$ & Second-order phase transition & First-order phase transition \\
 \colrule
 $1$ & $\times^{\text a}$ & $\times^{\text a}$\\
 $2$ & $q \le 4$ & $q > 4$ \\
 $d\ge 3$ & $q=1,2^{\text b}$ & $q \ge 3^{\text b}$ \\
 \botrule
\end{tabular}
}
 \begin{tabnote}
  $^{\text a}$ On one-dimensional lattice, phase transition does not occur.
  $^{\text b}$ There is no exact result of boundary value of $q$ between second-order phase transition and first-order phase transition. It is true that $q$-state Potts model on $d$-dimensional ($d\ge 3$) lattice for $q=1,2$ exhibits a second-order phase transition whereas a first-order phase transition takes place in that model for $q\ge 3$.
 \end{tabnote}
 \label{RTtable:sPotts}
\end{table}

\section{Preceding Models}

In this section we review two famous preceding models.
Both of two models are some kind of generalized Ising model.

\subsection{Blume-Capel Model}

We explain nature of the phase transition of the Blume-Capel model\cite{RTBlume-1966,RTCapel-1966}.
The Hamiltonian of this model is given as
\begin{eqnarray}
 {\cal H}_{\rm BC} = -J \sum_{\langle i,j \rangle} t_i t_j - D \sum_i t_i^2,
  \quad (J>0),
  \quad
  t_i = +1, 0, -1.
\end{eqnarray}
Here we refer to $t_i=0$ as vacancy.
At $D=-\infty$, vacancy is suppressed and the state of each spin is $t_i=+1$ or $-1$, 
whereas at $D=+\infty$, all spins become vacancies.
This model can represent magnetic lattice gas or annealed diluted Ising model, where $D$ corresponds to the chemical potential.
This model has been widely used analysis of multicritical phenomena in metallic alloys and liquid crystals {\it etc}\cite{RTCladis-1977,RTBlatter-1985,RTZhang-1995,RTRastogi-1999}.

\begin{figure}[b]
 \begin{center}
  \psfig{file=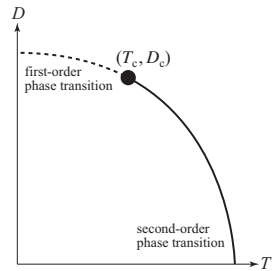,scale=0.75}
 \end{center}
 \caption{
 Schematic phase diagram of the Blume-Capel model.
 The solid and dotted curves represent second-order phase transition and first-order phase transition, respectively.
 The circle indicates the tricritical point $(T_{\rm c}, D_{\rm c})$.
 }
 \label{RTfig:BCPT}
\end{figure}

Schematic phase diagram of the Blume-Capel model is depicted in Fig.~\ref{RTfig:BCPT}.
The solid and dotted curves are second-order phase transition and first-order phase transition, respectively.
The circle between the solid and dotted curves in Fig.~\ref{RTfig:BCPT} represents the tricritical point ($T_{\rm c}$, $D_{\rm c}$).
Thus the Blume-Capel model is a fundamental model which has a tricritical point.
In this model we can change the order of the phase transition by controlling the chemical potential $D$.
When we decrease temperature for fixed $D(<D_{\rm c})$, a second-order phase transition occurs, whereas a first-order phase transition takes place when we decrease temperature for fixed $D(>D_{\rm c})$.
It should be noted that the ground state is also changed when we control $D$.
Then this type of control method for changing the order of the phase transition is not suitable for annealing methods.

\subsection{Wajnflasz-Pick Model}

Next we review nature of the phase transition of the Wajnflasz-Pick model\cite{RTWajnflasz-1971}.
The Hamiltonian of the Wajnflasz-Pick model is given as
\begin{eqnarray}
 {\cal H}_{\rm WP} = -J \sum_{\langle i,j \rangle} s_i s_j - h \sum_i s_i,
  \quad 
  s_i = \underbrace{+1, \cdots, +1}_{g_+}, \underbrace{-1,\cdots,-1}_{g_-},
\end{eqnarray}
where $g_+$ and $g_-$ are the number of $+1$-state and that of $-1$-state, respectively, and we assume $J>0$.
Note that this model for $g_+=g_-=1$ corresponds to the standard Ising model.
We can transform this Hamiltonian into the following Hamiltonian at finite temperature $T$:
\begin{eqnarray}
 \label{RTeq:WPt}
 {\cal H}_{\rm WP}' = -J \sum_{\langle i,j\rangle} \sigma_i^z \sigma_j^z 
  - (h-\frac{T}{2}\log \frac{g_+}{g_-}) \sum_i \sigma_i^z,
  \quad
  \sigma_i^z = +1, -1.
\end{eqnarray}
Note that the number of $+1$-state and that of $-1$-state are unity.
These two Hamiltonians are equivalent since the partition functions of these Hamiltonians are the same, {\it i.e.} ${\rm Tr}\, {\rm e}^{-\beta {\cal H}_{\rm WP}} = {\rm Tr}\, {\rm e}^{-\beta {\cal H}_{\rm WP}'}$.
The second term of Eq.~(\ref{RTeq:WPt}) consists of the original external magnetic field $h$ and temperature-dependent part.
The latter comes from the entropy effect of the bias of $g_+$ and $g_-$.
When $g_+=g_-$, this term disappears, and the Hamiltonian given by Eq.~(\ref{RTeq:WPt}) becomes the standard Ising model.
From this fact, the temperature-dependent external magnetic field can be regarded as the entropy-induced internal field.
The Wajnflasz-Pick model has been adopted for analysis of phase transition in spin-crossover materials\cite{RTZimmermann-1983,RTBoukheddaden-2000,RTMiyashita-2003,RTNishino-2003,RTTokoro-2006}.

\begin{figure}[t]
 \begin{center}
  \psfig{file=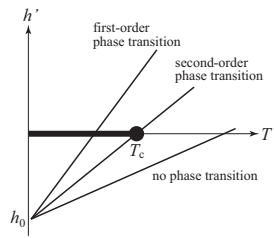,scale=0.75}
 \end{center}
 \caption{
 The bold line and the circle represent ferromagnetic phase and the critical point $T_{\rm c}$ of the Ising model.
 The thin lines correspond to changing the temperature at fixed external magnetic field $h_0$.
 The gradients of these lines are equal to $\log (g_+/g_-)/2$.
 }
 \label{RTfig:WPPT}
\end{figure}

We can analyze phase transition in this model by using the phase diagram of the standard Ising model.
Here, the coefficient of second term in Eq.~(\ref{RTeq:WPt}) is represented as
\begin{eqnarray}
 \label{RTeq:WPheff}
h' := h - \frac{T}{2} \log \frac{g_+}{g_-}.
\end{eqnarray}
Then the Hamiltonian given by Eq. (\ref{RTeq:WPt}) becomes the standard Ising model with external magnetic field $h'$, and the $h'-T$ phase diagram of this model is shown in Fig.~\ref{RTfig:WPPT}.
We consider the case for fixed finite external magnetic field $h=h_0$.
If we change the ratio $g_+/g_-$, we can control the order of the phase transition as depicted in Fig.~\ref{RTfig:WPPT}.
In this model, to change temperature corresponds to the tilted lines in Fig.~\ref{RTfig:WPPT},
and a slope of trajectory changes by the ratio $g_+/g_-$ according to Eq.~(\ref{RTeq:WPheff}).
If we set the ratio $g_+/g_- = \exp(2h_0/T_{\rm c}) =:g^*$, where $T_{\rm c}$ is the critical point of the standard Ising model, a second-order phase transition occurs when we decrease temperature.
If the ratio $g_+/g_-$ is smaller than $g^*$, no phase transition occurs, whereas a first-order phase transition takes place when $g_+/g_-$ is larger than $g^*$.
Thus, the Wajnflasz-Pick model is a standard model where we can change the order of the phase transition without changing the ground state by just controlling the ratio $g_+/g_-$.
It should be noted that the all phase transitions in this model do not accompany the twofold symmetry breaking which is the characteristic property of the standard Ising model, except for $h_0=0$ and $g_+=g_-$
({\it i.e.} the standard ferromagnetic Ising model without external field).

\section{Potts Model with Invisible States}

In Section 2, we reviewed nature of the phase transition in the standard ferromagnetic Potts model\cite{RTPotts-1952,RTWu-1982}.
It has been considered that the relation between the order of the phase transition and the symmetry which breaks at the transition point was investigated completely.
The standard ferromagnetic Potts model has actually succeeded to analyze a phase transition with discrete symmetry breaking appeared in real materials and complicated theoretical models.
Very recently, however, some phase transitions which are not consistent with the nature of the phase transition in the standard Potts model were reported, although these phase transitions accompany discrete symmetry breaking\cite{RTTamura-2008,RTStoudenmire-2009,RTOkumura-2010,RTTamura-2011}.
For instance, a first-order phase transition with threefold symmetry breaking occurs on two-dimensional lattice.
According to Table \ref{RTtable:sPotts}, the 3-state ferromagnetic Potts model on two-dimensional lattice should exhibit a second-order phase transition with threefold symmetry breaking.
This behavior is controversial feature.
In order to understand what happens in such phase transitions, we should propose a new model.

As we mentioned in Section 3, we can control the order of the phase transition by changing the chemical potential of vacancy or changing the bias of the number of states.
Roughly speaking, the Blume-Capel model and the Wajnflasz-Pick model change the internal energy and the entropy, respectively.
Motivated by these models, we constructed a new model -- Potts model with invisible states\cite{RTTamura-2010,RTTanaka-2011d,RTTanaka-2011e} to explain the nature of the above mentioned intriguing phase transition.
The Hamiltonian of this model is given as
\begin{eqnarray}
 \label{RTeq:PI}
 {\cal H}_{\rm PI} = -J \sum_{\langle i,j \rangle} \delta_{t_i,t_j} \sum_{\alpha=1}^q \delta_{t_i,\alpha},
  \quad
  t_i = 1, \cdots, q, q+1, \cdots, q+r,
\end{eqnarray}
where the states $1 \le t_i \le q$ and $q+1 \le t_i \le q+r$ are referred to as colored states and invisible states, respectively, and we assume $J>0$.
Hereafter we refer to this model as ($q$,$r$)-state Potts model.
Obviously, this model for $r=0$ corresponds to the standard Potts model,
and then,
this model is regarded as the straight forward extension of the standard Potts model.

\begin{figure}[b]
 \begin{center}
  \psfig{file=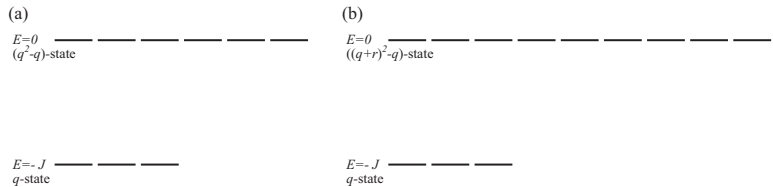,scale=0.75}
 \end{center}
 \caption{
 (a) Energy structure of $q$-state Potts model for $N=2$.
 The number of ground states and that of excited states are $q$ and $q^2-q$, respectively.
 (b) Energy structure of ($q$,$r$)-state Potts model for $N=2$.
 The number of ground states and that of excited states are $q$ and $(q+r)^2-q$, respectively.
 }
 \label{RTfig:twospin-ed}
\end{figure}

First, we consider two spin case for simplicity {\it i.e.} ${\cal H}_{\rm PI}=-J \delta_{t_1,t_2} \sum_{\alpha=1}^q \delta_{t_1,\alpha}$.
When $1\le t_1=t_2 \le q$, the energy becomes $-J$, otherwise the energy becomes zero.
Notice that even when $t_1 = t_2$, the energy becomes zero if $q+1 \le t_1=t_2 \le q+r$, which differs from the case of the standard ferromagnetic Potts model.
The energy structure of this model is depicted in Fig.~\ref{RTfig:twospin-ed}.
The number of ground states is $q$ and the number of excited states is $(q+r)^2-q$, whereas the number of excited states is $q^2-q$ for $r=0$ (standard Potts model).
Only the number of excited states increases by adding the invisible states into the standard Potts model.
It should be noted that the number of ground states of the model given by Eq.~(\ref{RTeq:PI}) and that of the standard Potts model are the same.

At finite temperature $T$, we can transform this Hamiltonian into the following form as well as the Wajnflasz-Pick model:
\begin{eqnarray}
  \label{RTeq:PIt}
 {\cal H}_{\rm PI}' = -J \sum_{\langle i,j \rangle} \delta_{s_i,s_j} \sum_{\alpha=1}^q \delta_{s_i,\alpha} - T\log r \sum_{i} \delta_{s_i,0},
  \quad
  s_i = 0, 1, \cdots, q.
\end{eqnarray}
Here we rename the label of invisible states from $q+1 \le t_i \le q+r$ to $s_i=0$.
The partition functions of both Hamiltonians are the same: ${\rm Tr}\, {\rm e}^{-\beta {\cal H}_{\rm PI}} = {\rm Tr}\, {\rm e}^{-\beta {\cal H}_{\rm PI}'}$.
The second term of this Hamiltonian is regarded as the chemical potential of invisible states and comes from the entropy effect of the number of invisible states $r$.

\subsection{Mean-Field Analysis}

\begin{figure}[t]
 \begin{center}
  $\begin{array}{ccc}
  \psfig{file=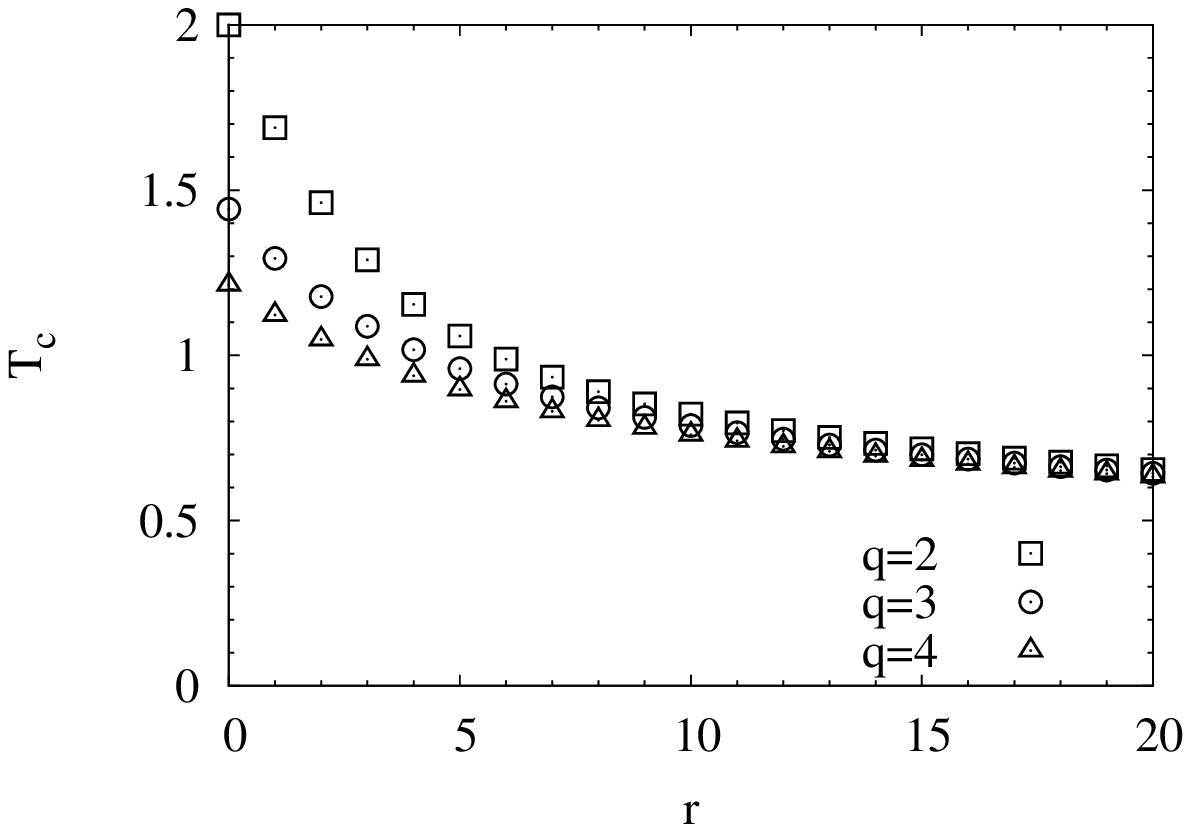,scale=0.45}&
   \hspace{5mm}&
  \psfig{file=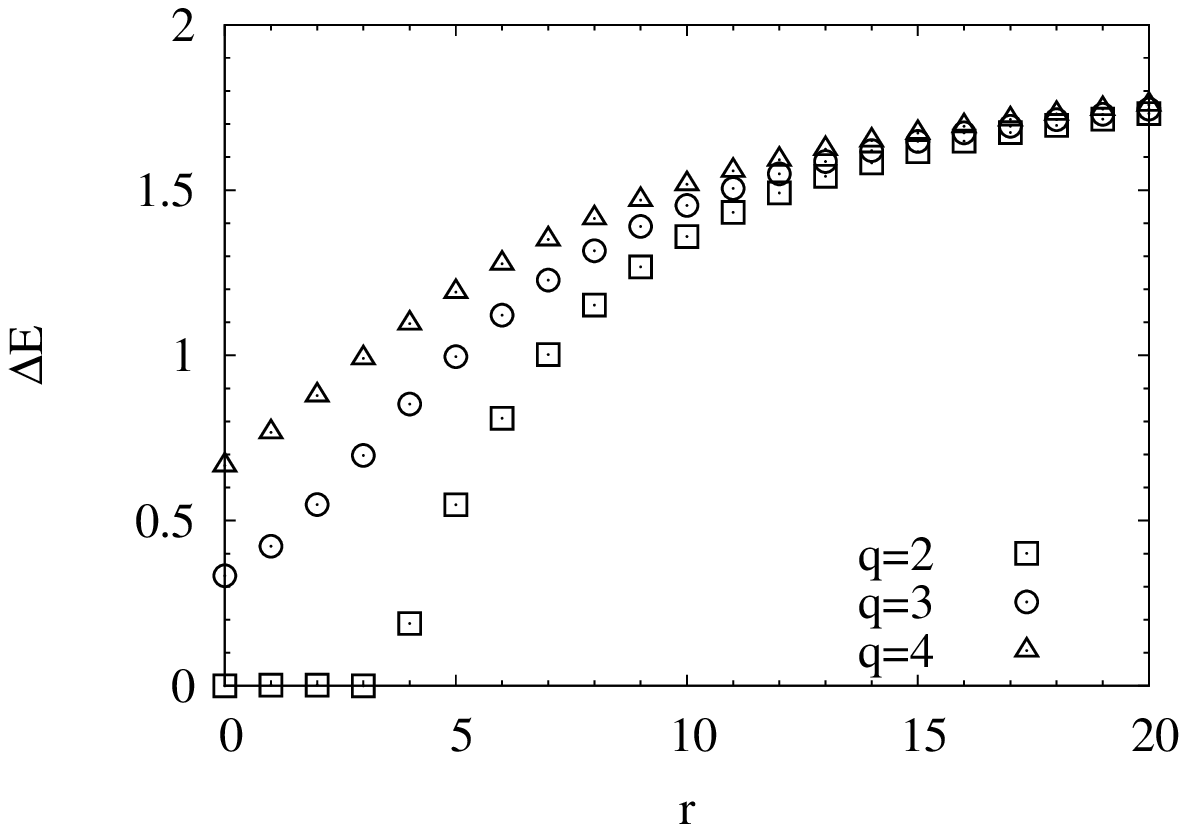,scale=0.45}\\
    ({\rm a}) & \hspace{5mm} & ({\rm b})
  \end{array}$
 \end{center}
 \caption{
 (a) Transition temperature as a function of the number of invisible states $r$ for $z=4$.
 (b) Latent heat as a function of $r$ for $z=4$.
 Both results are obtained by the Bragg-Williams approximation.
 }
 \label{RTfig:MFTc}
\end{figure}

We study phase transition in the ($q$,$r$)-state Potts model by the Bragg-Williams approximation as well as the standard ferromagnetic $q$-state Potts model\cite{RTKihara-1954}.
Since the number of ground states of this model is $q$, if a phase transition takes place, $q$-fold symmetry breaks at the transition point $T_{\rm c}$.
Let $w_\alpha$ be the fraction of the $\alpha$-th state ($0 \le \alpha \le q$).
Here we use the notation of the Hamiltonian given by Eq.~(\ref{RTeq:PIt}) and then the label of the invisible state is zero.
We assume that the first state is selected below the transition point $T_{\rm c}$ after $q$-fold symmetry breaks.
Then $\{w_\alpha\}$ can be represented as
\begin{eqnarray}
 \begin{cases}
  w_0 = y\\
  w_1 = \frac{1}{q}(1-y)[1+(q-1)x]\\
  w_\alpha = \frac{1}{q}(1-y)(1-x)
  \qquad \qquad (2 \le \alpha \le q)
 \end{cases},
\end{eqnarray}
where $0 \le x, y \le 1$. The fraction of the invisible states is represented by $y$, and $x$ indicates the degree of ordering.
When $x=0$, the system is paramagnetic state, whereas when $x=1$, the system is completely ordered state.
It is a natural condition that $w_\alpha$'s for $2 \le \alpha \le q$ are the same.
The internal energy and the entropy in the level of the Bragg-Williams approximation are given by
\begin{eqnarray}
 \label{RTeq:EBW}
 &&E^{\rm BW}(x,y) = -w_0 T \log r - \frac{zJ}{2} \sum_{\alpha=1}^q w_\alpha^2,\\
 &&S^{\rm BW}(x,y) = - \sum_{\alpha=0}^q w_\alpha \log w_\alpha,
\end{eqnarray}
respectively. The parameter $z$ in Eq.~(\ref{RTeq:EBW}) is the number of the nearest-neighbor sites.
As a result, the free energy can be written as
\begin{eqnarray}
 \nonumber
 &&F^{\rm BW}(x,y) = E^{\rm BW}(x,y) - TS^{\rm BW}(x,y)\\
 \nonumber
 &&= - \frac{zJ(1-y)^2}{2q}[(q-1)x^2+1]+yT\log \frac{y}{r}\\
 && + (1-y)T 
  \left[
   \frac{1+(q-1)x}{q}\log \frac{1+(q-1)x}{1-x} + \log \frac{(1-y)(1-x)}{q}
  \right].
\end{eqnarray}
From this free energy, the transition temperature and the latent heat can be obtained by numerical calculation.
The number of invisible states $r$ dependencies of the transition temperature and the latent heat for $q=2,3,$ and $4$ are shown in Fig.~\ref{RTfig:MFTc}.

The latent heat expresses the energy difference between metastable state and stable state at the transition temperature.
By the definition, if the latent heat is a finite value, the internal energy jumps at the transition temperature.
Thus a finite latent heat indicates that a first-order phase transition occurs.
From Fig.~\ref{RTfig:MFTc} (b), the ($2$,$r$)-state Potts model for $r \le 3$ exhibits a second-order phase transition and others have a first-order phase transition.
As the number of invisible states $r$ increases, the transition temperature decreases and the latent heat increases.
This result is quite natural since the invisible states contribute to the entropy.
To clarify the effect of the invisible states, we compare with the transition temperature and the latent heat for $r=0$ (standard Potts model).
The transition temperature and the latent heat for $r=0$ are obtained as
\begin{eqnarray}
 &&\begin{cases}
  T_{\rm c}^{\rm BW} (q=2,r=0) = \frac{zJ}{2}\\
  \Delta E^{\rm BW} (q=2,r=0) = 0
 \end{cases},
\\
 &&\begin{cases}
  T_{\rm c}^{\rm BW} (q\ge 3,r=0) = \frac{zJ}{2\log(q-1)}\left( \frac{q-2}{q-1}\right)\\
  \Delta E^{\rm BW} (q\ge 3,r=0) = \frac{zJ(q-2)^2}{2q(q-1)}
 \end{cases}.
\end{eqnarray}
The transition temperature and the latent heat for $r=0$ and $z=4$ are also shown in Fig.~\ref{RTfig:MFTc}.

We also consider the relation between the ($2$,$r$)-state Potts model and the Blume-Emery-Griffiths model\cite{RTBlume-1971} which is an extended model of the Blume-Capel model.
The Hamiltonian of the Blume-Emery-Griffiths model is given by
\begin{eqnarray}
 {\cal H}_{\rm BEG} = -\frac{J}{2} \sum_{\langle i,j \rangle} t_i t_j
  -\frac{J'}{2} \sum_{\langle i,j \rangle} t_i^2 t_j^2 - D \sum_i (1-t_i^2),
  \quad
  t_i = +1, 0, -1,
\end{eqnarray}
where the biquadratic interaction (the second term) is introduced into the Blume-Capel model.
The Blume-Emery-Griffiths model for $J=J'$ and $D=T\log r$ is equivalent to the ($2$,$r$)-state Potts model.
In order to investigate nature of the phase transition of this model, we obtain the phase diagram of the model for $J=J'$ and unfixed $D$ by the Bragg-Williams approximation for $z=4$.

\begin{figure}[t]
 \begin{center}
  \psfig{file=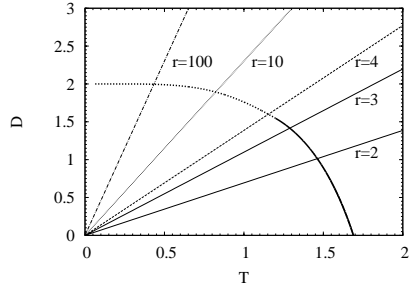,scale=0.45}
 \end{center}
 \caption{
 Phase diagram of the Blume-Emery-Griffiths model for $J=J'$ obtained by the Bragg-Williams approximation for $z=4$.
 The thick solid and dotted curves indicate second-order phase transition and first-order phase transition.
 The thin lines correspond to decreasing temperature for $r=2$, $3$, $4$, $10$, and $100$ from bottom to top.
 The gradients of these lines are $\log r$.
 }
 \label{RTfig:BEG}
\end{figure}

The phase diagram of the Blume-Emery-Griffiths model for $J=J'$ is depicted in Fig.~\ref{RTfig:BEG}.
The thick solid and dotted curves indicate second-order phase transition and first-order phase transition, respectively.
Here we consider the case for $D=T\log r$.
To change temperature corresponds to the thin lines in Fig.~\ref{RTfig:BEG}.
The gradients of these lines are $\log r$.
When we decrease temperature for the case of $r=2$ and $3$, the lines cross a second-order phase transition curve whereas the lines for large $r$ cross a first-order phase transition curve.
This result is consistent with the result shown in Fig.~\ref{RTfig:MFTc}.

\subsection{Monte Carlo Simulation}

In the previous subsection we studied the order of the phase transition and the latent heat in the ($q$,$r$)-state Potts model by the Bragg-Williams approximation.
These results correspond to the infinite-dimensional version of the ($q$,$r$)-state Potts model.
Then in order to consider nature of the phase transition in the ($q$,$r$)-state Potts model on finite-dimensional lattice, we study thermodynamic properties of this model on the $L\times L(=N)$ square lattice by Monte Carlo simulation.

First we calculate temperature dependencies of the order parameter ${\bf m}_{\rm Potts}$, the density of invisible states $\rho_{\rm inv}$, the internal energy $E$, and the specific heat $C$ for ($q$,$r$)=($4$,$20$).
The definition of $\rho_{\rm inv}$ is defined by
\begin{eqnarray}
 \rho_{\rm inv} = \frac{1}{N} \sum_i \sum_{\alpha=q+1}^{q+r} \delta_{t_i,\alpha}.
\end{eqnarray}
It should be noted that we use the order parameter of the standard ferromagnetic $q$-state Potts model given by Eq.~(\ref{RTeq:ORDPotts}).
\begin{figure}[t]
 \begin{center}
  $\begin{array}{ccc}
   \psfig{file=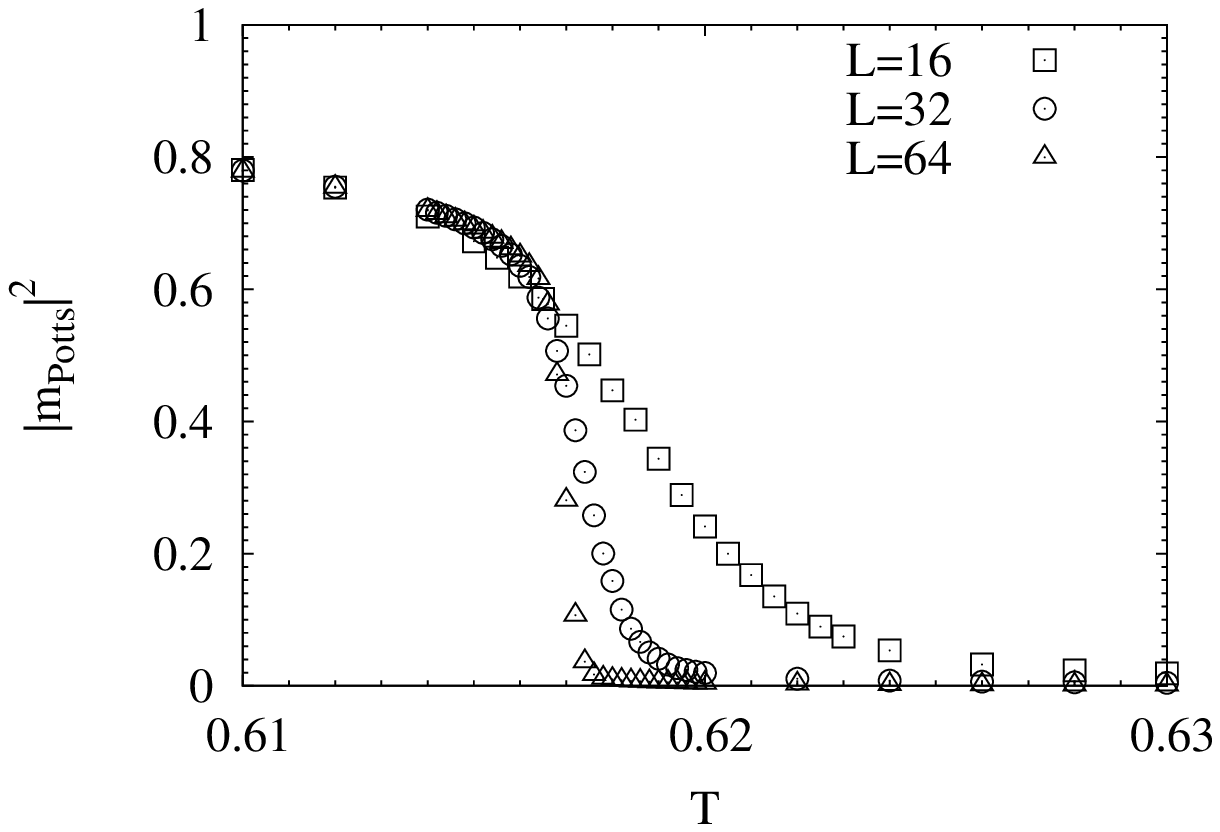,scale=0.45} & \hspace{5mm} &
    \psfig{file=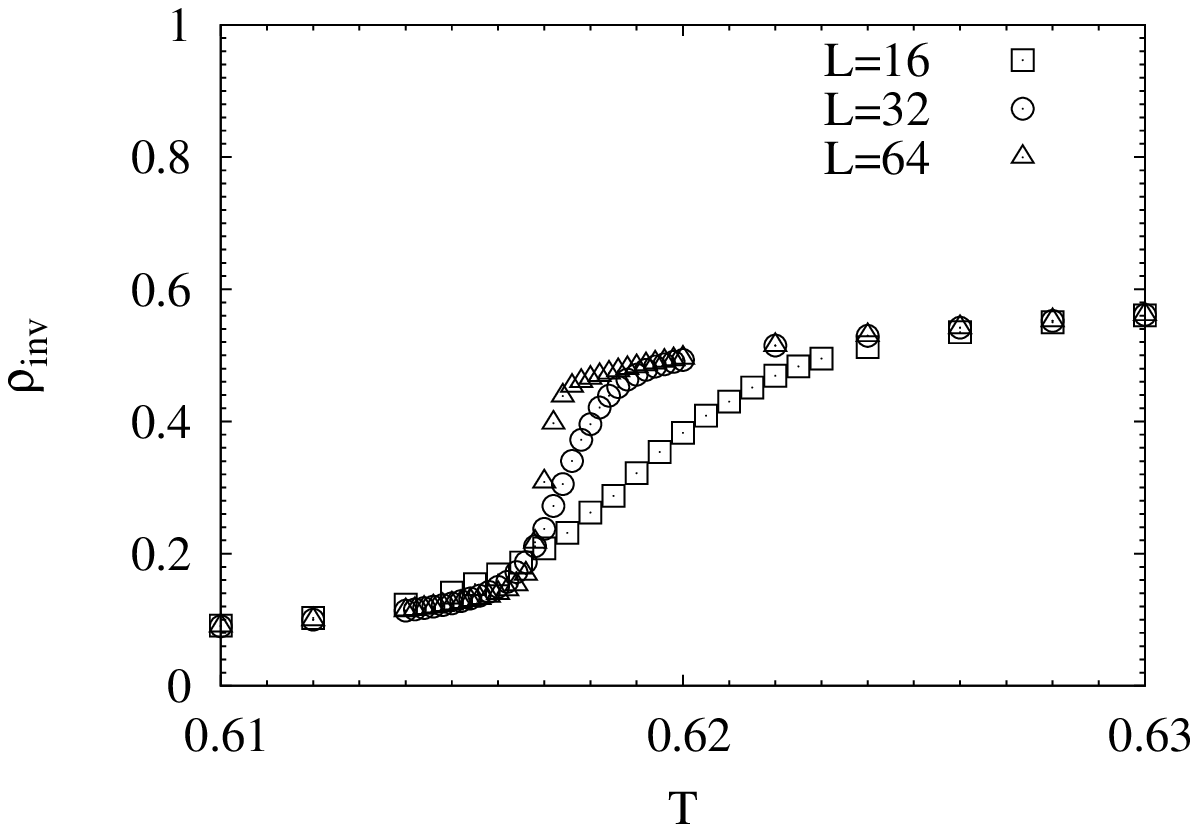,scale=0.45}\\
    ({\rm a}) & \hspace{5mm} & ({\rm b})\\
    \psfig{file=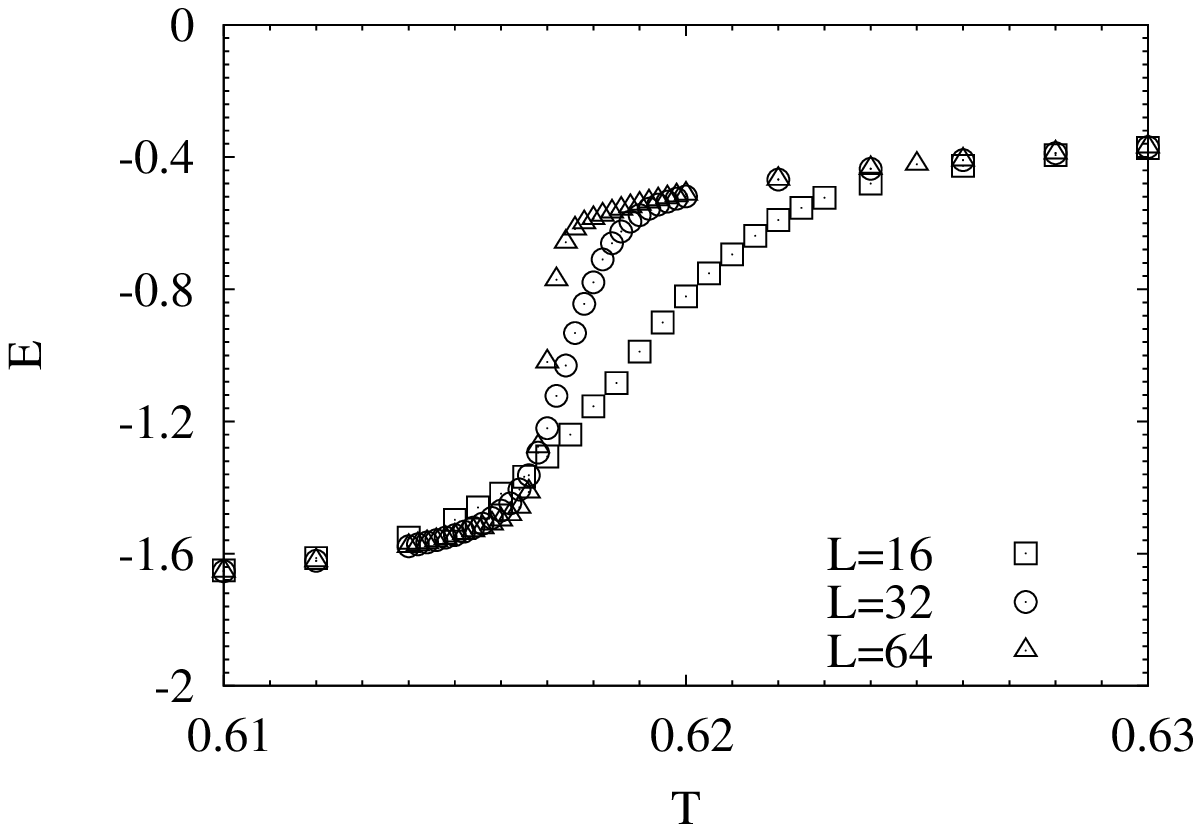,scale=0.45} & \hspace{5mm} &
    \psfig{file=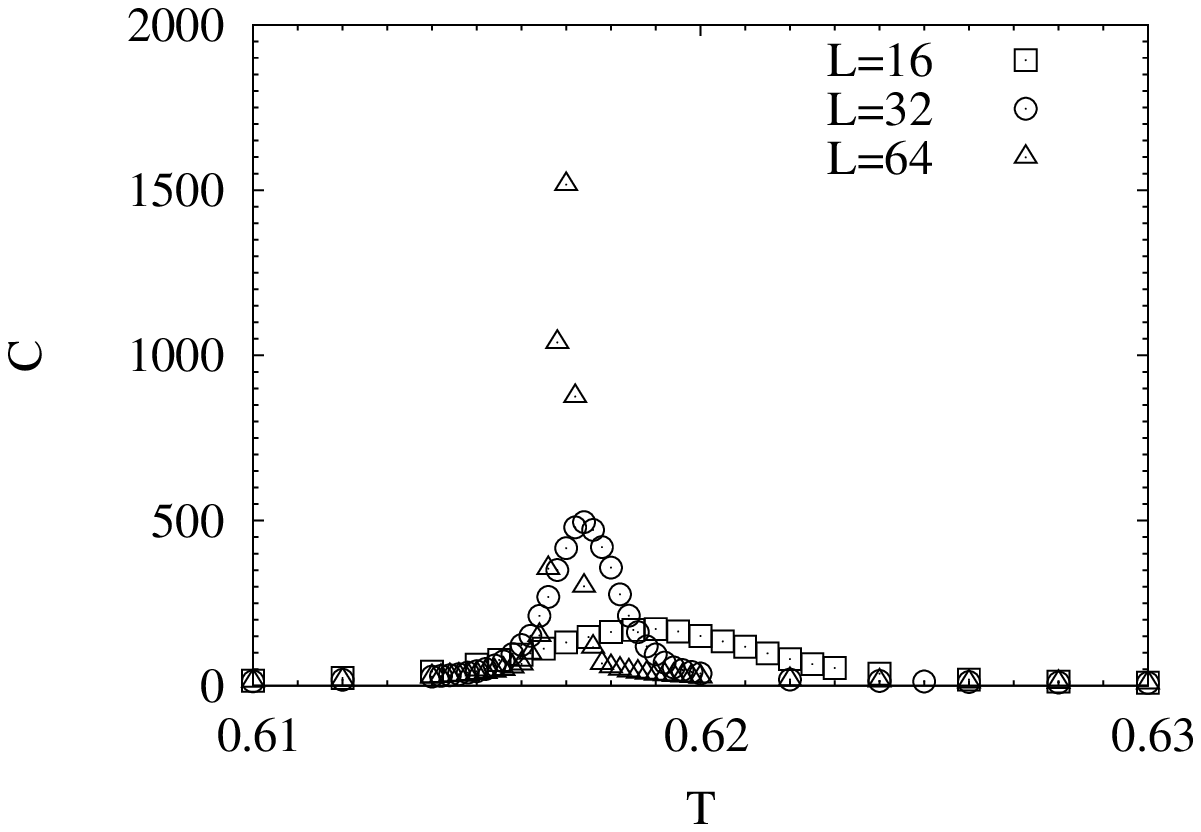,scale=0.45}\\
    ({\rm c}) & \hspace{5mm} & ({\rm d})
   \end{array}$
 \end{center}
 \caption{
 Temperature dependencies of (a) the order parameter, (b) the fraction of the invisible states, (c) the internal energy, and (d) the specific heat for the ($4$,$20$)-state Potts model.
 The error bars are omitted for clarity since they are smaller than the symbol size.
 }
 \label{RTfig:MC-PQ}
\end{figure}
These quantities as functions of temperature are shown in Fig.~\ref{RTfig:MC-PQ}.
Figure \ref{RTfig:MC-PQ} (a) shows temperature dependency of the order parameter $|{\bf m}_{\rm Potts}|^2$, which indicates fourfold symmetry breaks at the transition temperature.
The fraction of the invisible states as a function of temperature is shown in Fig.~\ref{RTfig:MC-PQ} (b).
As the temperature increases, the fraction of invisible states increases.
At $T=\infty$, this value becomes $r/(q+r) = 0.833 \cdots$ by the definition whereas this value becomes zero at $T=0$.
Figure \ref{RTfig:MC-PQ} (c) and (d) display the internal energy and the specific heat, respectively.
The specific heat is not divergent behavior but has a finite large value at the transition temperature.
This result indicates that a first-order phase transition takes place.

In order to confirm that this phase transition is of first-order, we calculate energy histogram at the temperature $T_{\rm c}(L)$ where the specific heat has the maximum value.
The temperature $T_{\rm c}(L)$ is obtained by reweighting method\cite{RTFerrenberg-1988,RTFerrenberg-1989}.
The energy histogram $P(E)$ is defined as
\begin{eqnarray}
 P(E) = W(E) {\rm e}^{-\beta E},
\end{eqnarray}
where $W(E)$ is the number of states in energy $E$.
\begin{figure}[t]
 \begin{center}
  \psfig{file=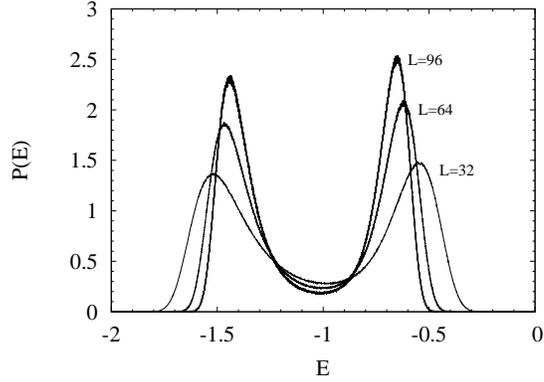,scale=0.60}
 \end{center}
 \caption{
 The energy histogram at the temperature $T_{\rm c}(L)$ for the ($4$,$20$)-state Potts model.
 The temperatures are used the following values $T_{\rm c}(L=32) = 0.61740$, $T_{\rm c}(L=64)=0.61698$, and $T_{\rm c}(L=96)=0.616894$, respectively.
 }
 \label{RTfig:bimodal}
\end{figure}
Figure \ref{RTfig:bimodal} shows the energy histogram at $T_{\rm c}(L)$ for $L=32$, $64$, and $96$.
A bimodal distribution is observed, which is a characteristic behavior of system which exhibits a first-order phase transition.
As the number of spins increases, the two peaks become sharp.
In the thermodynamic limit, these two peaks are expected to be the delta functions.

Next we take finite-size scaling in order to determine the transition temperature and the latent heat in the thermodynamic limit.
We adopt the following functions\cite{RTChalla-1986}:
\begin{eqnarray}
 \label{RTeq:fitting1}
 &T_{\rm c}(L) = a L^{-d} + T_{\rm c},\\
 \label{RTeq:fitting2}
 &C_{\rm max}(L) \propto \frac{(\Delta E)^2 L^d}{4T_{\rm c}^2},
\end{eqnarray}
where $C_{\rm max}(L)$ and $\Delta E$ are the maximum value of the specific heat for $L\times L$ system and the latent heat in the thermodynamic limit, respectively.
\begin{figure}[t]
 \begin{center}
  $\begin{array}{ccc}
  \psfig{file=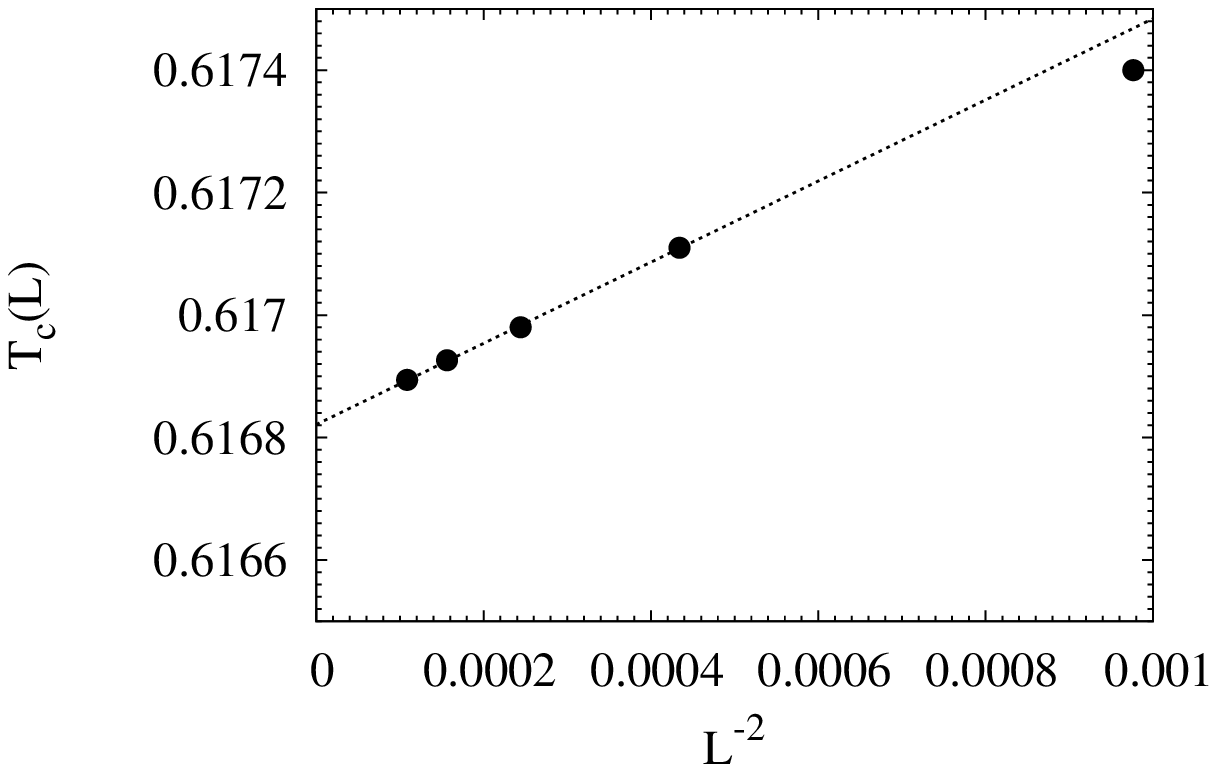,scale=0.4}&\hspace{5mm}&
  \psfig{file=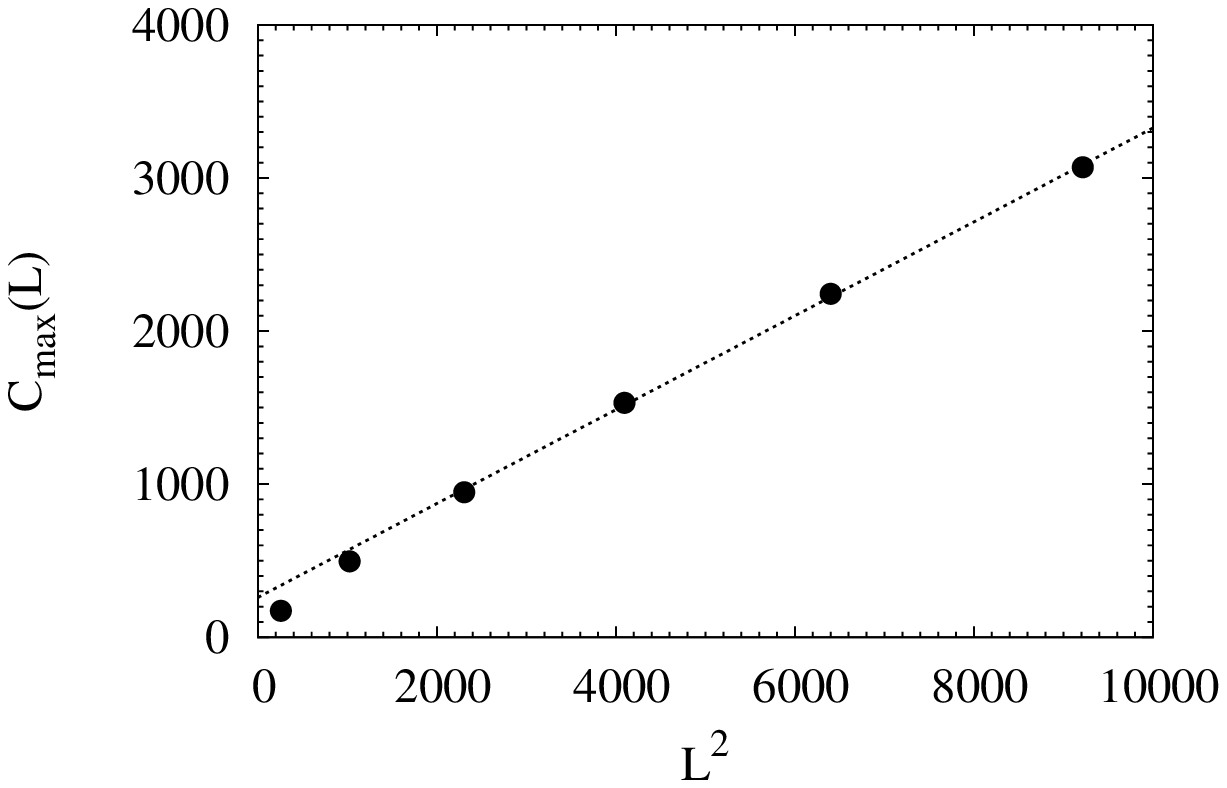,scale=0.4}\\
    ({\rm a}) & \hspace{5mm}& ({\rm b})
  \end{array}$
 \end{center}
 \caption{
 (a) The temperature $T_{\rm c}(L)$ as a function of $L^{-2}$ for the ($4$,$20$)-state Potts model.
 The dotted line is a fitting line given by Eq.~(\ref{RTeq:fitting1}).
 The intercept of the fitting line indicates the transition point $T_{\rm c}$.
 (b) The maximum value of the specific heat as a function of $L^2$ for the ($4$,$20$)-state Potts model.
 The dotted line is a fitting line given by Eq.~(\ref{RTeq:fitting2}).
 }
 \label{RTfig:scaling}
\end{figure}
In this case, we adopt $d=2$.
Figure \ref{RTfig:scaling} (a) shows $T_{\rm c}(L)$ as a function of $L^{-2}$.
The intercept of the dotted line corresponds to the transition temperature $T_{\rm c}$.
The maximum value of the specific heat $C_{\rm max}(L)$ as a function of $L^2$ is depicted in Fig.~\ref{RTfig:scaling} (b).
From these results, the transition temperature and the latent heat are obtained as $T_{\rm c}=0.61683(1)$ and $\Delta E=0.68(2)$, respectively.
This fact indicates that we succeeded to construct model which exhibits a first-order phase transition with fourfold symmetry breaking on the two-dimensional lattice.
By the same procedure, we can obtain the transition temperature and the latent heat for several ($q$,$r$) (see Table \ref{RTtable:qrPotts}).

\begin{table}[t]
\tbl{The transition temperature and the latent heat for several ($q$,$r$) on the square lattice.}
{
\begin{tabular}{ccc}\toprule
 ($q,r$) & Latent heat & Transition temperature \\
 \colrule
 ($2$,$0$)$^{\text a}$  & $0$       & $1.13459$    \\
 ($2$,$30$)$^{\text b}$ & $1.02(2)$ & $0.57837(1)$ \\
 ($2$,$32$)$^{\text c}$ & $1.23(2)$ & $0.56857(1)$ \\
 \colrule
 ($3$,$0$)$^{\text a}$  & $0$       & $0.994973$   \\
 ($3$,$25$)$^{\text b}$ & $0.81(2)$ & $0.59630(1)$ \\
 ($3$,$27$)$^{\text c}$ & $1.05(2)$ & $0.58513(1)$ \\
 \colrule
 ($4$,$0$)$^{\text a}$  & $0$       & $0.910239$   \\
 ($4$,$20$)$^{\text b}$ & $0.68(2)$ & $0.61683(1)$ \\
 ($4$,$22$)$^{\text c}$ & $0.87(2)$ & $0.60396(1)$ \\
 \botrule
\end{tabular}
}
 \begin{tabnote}
  $^{\text a}$ These results were obtained exactly.
  $^{\text b}$ These results were first obtained in Ref.~\refcite{RTTamura-2010}.
  $^{\text c}$ These results were obtained in Ref.~\refcite{RTTanaka-2011d}.
 \end{tabnote}
 \label{RTtable:qrPotts}
\end{table}

As the number of invisible states increases, the transition temperature decreases and the latent heat increases, which is qualitatively consistent with the result obtained by the Bragg-Williams approximation.
Thus, we conclude that invisible state in the Potts model stimulates a first-order phase transition.
From this, unfortunately, this model is not useful for the optimization problems.
Recently, nature of the phase transition in the ($q$,$r$)-state Potts model has been confirmed by a number of researchers\cite{RTEnter-2011a,RTEnter-2011b,RTMori-2011}.

\subsection{Another Representation of the Potts Model with Invisible States}

So far, we showed thermodynamic properties and phase transition of the ($q$,$r$)-state Potts model.
This model is regarded as a straightforward extension of the standard $q$-state Potts model.
In order to make it more clear, we show another representation of the Potts model with invisible states.
First we consider the standard $q$-state Potts model.
Let $\vec{S}_i$ be a state vector at the $i$-th site.
The state vector is $q$-dimensional binary vector.
Only one element in this vector is one and the others are zero.
The position of one corresponds to the state.
For example, when $\vec{S_i}={}^{\rm T}(0,0,1,0,\cdots,0)$, the state of the $i$-th site is the third state.
Here $^{\rm T}\vec{v}$ represents the transpose of the vector $\vec{v}$.
We can express the Hamiltonian by matrix representation as follows:
\begin{eqnarray}
 {\cal H}_{\rm Potts} = -J \sum_{\langle i,j \rangle} \delta_{s_i,s_j} = 
  -\sum_{\langle i,j \rangle} {}^{\rm T}\vec{S_i} \hat{J}_{\rm Potts} \vec{S_j},
\end{eqnarray}
where $\hat{J}_{\rm Potts}$ is a $q$-by-$q$ diagonal matrix whose elements are expressed as
\begin{eqnarray}
 \hat{J}_{\rm Potts} = {\rm diag}(\underbrace{J,\cdots,J}_{q}).
\end{eqnarray}
Next we consider such a representation of the ($q$,$r$)-state Potts model.
Let $\vec{T_i}$ be a ($q+r$)-dimensional indicator vector.
The Hamiltonian of the Potts model with invisible states is represented as
\begin{eqnarray}
 {\cal H}_{\rm PI} = -J \sum_{\langle i,j\rangle} \delta_{t_i,t_j} \sum_{\alpha=1}^q \delta_{t_i,\alpha}
  = -\sum_{\langle i,j \rangle} {}^{\rm T}\vec{T_i} \hat{J}_{\rm PI} \vec{T_j},
\end{eqnarray}
where $\hat{J}_{\rm PI}$ is a $(q+r)$-by-$(q+r)$ diagonal matrix whose elements are expressed as
\begin{eqnarray}
 \hat{J}_{\rm PI} = {\rm diag}(\underbrace{J,\cdots,J}_{q},\underbrace{0,\cdots,0}_{r}).
\end{eqnarray}
We can construct more generalized model by using the matrix representation by introducing the off-diagonal elements like the clock model, for instance.
It is an important topic to control the order of the phase transition by changing the form of the interaction matrix $\hat{J}$.

\section{Conclusion and Future Perspective}

In this paper we reviewed nature of first-order and second-order phase transitions in general by taking the Ising model for instance.
We demonstrated how to change the order of the phase transition by using the Blume-Capel model and the Wajnflasz-Pick model.
In the Blume-Capel model, we can change the order of the phase transition by changing the chemical potential of vacancy.
In this model, the ground state changes when we control the chemical potential.
In the Wajnflasz-Pick model, on the other hand, by changing the bias of the number of states, the order of the phase transition can be changed without changing the ground state.
By introducing the effects in both models into the standard Potts model, we constructed a generalized Potts model -- Potts model with invisible states.
The invisible state corresponds to the vacancy as well as the Blume-Capel model and the number of the invisible states contributes to the entropy as well as the Wajnflasz-Pick model.
We can control the order of the phase transition by just changing the number of invisible states without variation of ground state.
Then as the number of invisible states increases, the transition temperature decreases and the latent heat increases.

To change the order of the phase transition is an important topic for not only statistical physics but also computational science and quantum information theory (especially, quantum annealing or quantum adiabatic computation).
Simulated annealing and quantum annealing which are based on (quantum) statistical physics do not work well in systems which exhibit a first-order phase transition.
If a second-order phase transition takes place, a situation becomes not so bad, but a kind of critical slowing down problem remains.
Phase transition in systems is obstacle in optimization problems whenever we use some kind of annealing procedures.
Then it is anticipated to propose a method to control the order of the phase transition or to erase the phase transition if possible.
Unfortunately, the invisible state in the Potts model stimulates a first-order phase transition and is not useful for annealing methods in the present stage.
However our proposed method to generalize models is quite general and simple.
Thus it is expected that how to erase the phase transition or how to change the order of the phase transition will be found in a similar way. 

\section*{Acknowledgement}

The authors are grateful to Jie Lou, Yoshiki Matsuda, Seiji Miyashita, Takashi Mori, Yohsuke Murase, Taro Nakada, Masayuki Ohzeki, Per Arne Rikvold, and Eric Vincent for their valuable comments.
R.T. is partly supported by Global COE Program ``the Physical Sciences Frontier'', MEXT, Japan.
S.T. is partly supported by Grand-in-Aid for JSPS Fellows (23-7601).
The present work is financially supported by MEXT Grant-in-Aid for Scientific Research (B) (22340111), and for Scientific Research on Priority Areas ``Novel States of Matter Induced by Frustration'' (19052004), and by Next Generation Supercomputing Project, Nanoscience Program, MEXT, Japan.
The computation in the present work was performed on computers at the Supercomputer Center, Institute for Solid State Physics, University of Tokyo. 

\bibliographystyle{ws-procs9x6}
\bibliography{ws-pro-sample}

\end{document}